\documentclass[twocolumn,english]{revtex4}
\usepackage[T1]{fontenc}
\usepackage[latin1]{inputenc}
\usepackage{amssymb}

\makeatletter

\newcommand{\half}{\mbox{$\textstyle \frac{1}{2}$}}

\newcommand{\sixt}{\mbox{$\textstyle \frac{1}{6}$}}
\newcommand{\myfrac}[2]{\mbox{$\textstyle \frac{#1}{#2}$}}

\usepackage{babel}
\makeatother
\begin{document}

\title{New bulk scalar field solutions in brane worlds}

\author{M. Parry}

\author{S. Pichler}

\affiliation{Theoretische Physik, Ludwig-Maximilians-Universit\"{a}t, Theresienstr.
37, 80333 M\"{u}nchen, GERMANY}

\begin{abstract}
We use nonlinear perturbation theory to obtain new solutions for brane
world models that incorporate a massive bulk scalar field. We then
consider tensor perturbations and show that Newtonian gravity is recovered
on the brane for both a light scalar field and for a bulk field with
large negative mass. This latter result points to the viability of
higher-derivative theories of gravity in the context of bulk extra
dimensions.
\end{abstract}
\maketitle

\section{Introduction}

One of the unsatisfying features of the Randall-Sundrum models has
always been the need to fine-tune the brane tension and the bulk cosmological
constant. Immediately after these models were proposed, many authors
considered adding bulk scalar fields in order to evade this problem.
It was hoped that the brane-bulk system would also be stabilized by
the presence of these fields. However, it has proved difficult to
find self-consistent analytic solutions to all but the simplest scenarios,
e.g. superpotential models \cite{tye,brax,wolfe}. Furthermore, the
order of complexity rises when one considers a fully time-dependent
scenario. For this reason, work has recently been concentrated on
numerical simulations \cite{code}.

On the other hand, the insight of Randall and Sundrum was that gravity
can still be effectively 4-dimensional despite the presence of a fifth,
spatial dimension. Therefore, if one complicates their simple models
by including a bulk scalar field, it is vital to ensure that gravity
has the usual Newtonian limit, at the very least. 

In this paper, we tackle analytically the case of brane worlds with
a massive bulk scalar field. We conduct two distinct perturbative
analyses. In the first, we utilize ideas in nonlinear perturbation
theory to obtain leading corrections to the background warp factor
and bulk field; it is the nature of the former that is crucial to
the localization of gravity on the brane. Then, we consider metric
and matter perturbations of the brane-bulk system, and compute the
Green's function in response to a point-mass on the brane. We show
that the Green's function has the required $1/r$-behavior in the
long-distance limit.

Although our presentation here is something of a mathematical proof-of-concept,
it sets the stage for a subsequent paper that will explore these issues
in the context of theories of higher-derivative gravity.

\section{Background solutions\label{sec:Background-solutions}}

We begin our analysis by considering the model of one Minkowski brane
in a bulk that contains a massive scalar field $\phi$. The scalar
field potential is $V=\half M^{2}\phi^{2}.$ The brane is coupled
to the scalar field via its tension; to be specific, we choose a Liouville
potential for the tension, $\lambda(\phi)=e^{-2\phi/\sqrt{3}}.$ In
fact, the explicit choice of $\lambda(\phi)$ is not so important
since we only ever need its value at $\phi(0).$ Our choice effectively
fixes the degree of freedom associated with $\lambda^{\prime}(\phi(0))$.

The line-element for the brane-bulk system may be written as\begin{equation}
ds^{2}=e^{2X(y)}\eta_{\mu\nu}dx^{\mu}dx^{\nu}-dy^{2},\label{eq:RSmetric}\end{equation}
with the brane at $y=0.$ The equations of motion that follow from
the Einstein field equations are\begin{eqnarray}
-3X_{yy}-6X_{y}^{2} & = & \half\phi_{y}^{2}+\half M^{2}\phi^{2},\\
6X_{y}^{2} & = & \half\phi_{y}^{2}-\half M^{2}\phi^{2},\label{eq:constraint}\\
\phi_{yy}+4X_{y}\phi_{y} & = & M^{2}\phi.\end{eqnarray}
Note that we will often use a subscript to denote a derivative when
there is no cause for confusion. The junction conditions at the brane
position are \begin{equation}
X_{y}(0)=-\sixt,\quad\phi_{y}(0)=-\myfrac{1}{\sqrt{3}},\end{equation}
where we have assumed $\mbox{$\mathbb{Z}_{2}$}$-symmetry---we evaluate
quantities as $y\downarrow0$---and used the fact that $\phi(0)=0,$
which follows from eq. (\ref{eq:constraint}). Without loss of generality,
we choose $X(0)=0.$ Finally, we point out that we have made all quantities
dimensionless by rescaling in terms of a characteristic length $\lambda^{-1}(0)M_{P}^{3}$
and characteristic scalar field value $M_{P}^{3/2},$ where $M_{P}$
is the 5-dimensional Planck mass.

\subsection{Small $M^{2}$}

We let $\epsilon=M^{2}$ be our perturbative parameter. Then, we introduce
a ``strained'' coordinate,\begin{equation}
x=(1+\mu\epsilon)^{-1}y,\end{equation}
 in place of $y.$ This technique, known as the Lindstedt-Poincar\'{e} method
\cite{ali}, can make a divergent perturbative expansion convergent
by an appropriate choice of $\mu.$

We now let $X=X_{0}(x)+\epsilon X_{1}(x)+\ldots$ and $\phi=\phi_{0}(x)+\epsilon\phi_{1}(x)+\ldots$,
and expand the equations of motion and junction conditions up to first
order in $\epsilon.$ The zeroth order equations yield \begin{equation}
X_{0}=\myfrac14\ln\left(1-\myfrac23x\right),\quad\phi_{0}=2\sqrt{3}X_{0}.\end{equation}
The obvious feature of the solution is the singularity at $x=\frac{3}{2}.$
It is easy to check that this is not a coordinate singularity; indeed,
such singularities appear to be generic in scalar field models \cite{tye,wolfe}.
We will discuss what to do about them in subsection \ref{sub:Shielding-the-singularity}.

The first order equations are much more involved and we omit the expressions
for $X_{1}$ and $\phi_{1}.$ We are not primarily interested in their
contribution to the overall solution at this stage, but rather in
what they imply about $\mu.$ To make this last point more transparent,
it is enough to know that the homogeneous solution to $X_{1}$ is\begin{equation}
X_{1}^{(h)}=A(\mu)+\frac{B(\mu)}{1-\myfrac23x},\end{equation}
where $A$ and $B$ are constants. The problem is that the latter
term will dominate $X_{0}:$ even for small $\epsilon,$ the warp
factor $e^{X}$ will pick up a term like $(1-\frac{2}{3}x)^{-3/4}.$
Such a term is sometimes called a secular term. The solution is to
choose $\mu$ so that $B(\mu)=0.$ It turns out that\begin{equation}
\mu=-\myfrac16,\end{equation}
and then\begin{equation}
X_{0}=\myfrac14\ln\left(1-\frac{\frac{2}{3}y}{1-\frac{1}{6}\epsilon}\right).\end{equation}
Note that, even at zeroth order, an effect of the perturbation is
seen. However, we now need the first order terms to ensure the junction
conditions are satisfied. That the singularity occurs at $y=\frac{3}{2}(1-\frac{1}{6}\epsilon),$
is borne out by numerical simulations.

\subsection{Large $-M^{2}$}

In this case, the convenient perturbative parameter is $\epsilon=|M|^{-1}.$
The Lindstedt-Poincar\'{e} method is not required here, but because
the $\epsilon\rightarrow0$ limit is problematic, it is also not obvious
what perturbative expansion will work. From numerical observations,
it is apparent that $X\sim\frac{1}{2}\ln(1-\frac{1}{3}y)$ but that
highly oscillatory terms arise in $X_{y}$ and $X_{yy}.$ Curiously,
the solution for $X$ is what one would obtain in the usual Randall-Sundrum
case with no bulk cosmological constant \emph{if} eq. (\ref{eq:constraint})
could be ignored. Therefore, the limit we are considering here can
be thought of as the best a scalar field can do to mimic $\Lambda=0.$
Note that, once again, there is a singularity at finite coordinate
distance from the brane.

Expansions for $X$ and $\phi$ that capture the oscillatory nature
of the solutions are as follows:\begin{eqnarray}
X & = & \myfrac12\ln\left(1\!-\!\myfrac13y\right)\!+\!\sum_{n=1}^{\infty}\epsilon^{2n}\left(1\!-\!\myfrac13y\right)^{-2n}E_{n}\left(\frac{y}{\epsilon}\right)\!,\\
\phi & = & \sum_{n=1}^{\infty}\epsilon^{2n-1}\left(1\!-\!\myfrac13y\right)^{-2n+1}F_{n}\left(\frac{y}{\epsilon}\right).\end{eqnarray}
In fact, for our purposes, the first term under the summation will
be sufficient. The important point, however, is that taking derivatives
of these terms makes them of lower order in $\epsilon.$ The solutions
of the equations resulting from substituting the above expansions
into the equations of motion and the junction conditions are\begin{equation}
E_{1}(u)=-\myfrac{1}{36}\sin^{2}u,\quad F_{1}(u)=-\myfrac{1}{\sqrt{3}}\sin u.\end{equation}

\subsection{Eliminating the singularity\label{sub:Shielding-the-singularity}}

As pointed out above, the presence of singularities in our solutions
is both generic and problematic. The standard method of dealing with
them is to add a ``regulator'' brane to the system. If the singularity
is at $y=y_{s}$, then placing the second brane at $y=y_{\star}<y_{s}$
will, due to $\mbox{$\mathbb{Z}_{2}$}$-symmetry, ``slice off'' the
section $y>y_{\star}$, which includes the singularity.

The insertion point $y_{\star}$ is not arbitrary, however. If we
let the brane tension of the regulator brane be $\lambda_{\star}(\phi),$
then a second set of junction conditions which must be satisfied are\begin{equation}
X_{y}(y_{\star})=\myfrac16\lambda_{\star}(\phi),\quad\phi_{y}(y_{\star})=-\half\lambda_{\star}^{\prime}(\phi),\end{equation}
where all quantities are evaluated as $y\uparrow y_{\star}.$ In what
follows, we will assume that these conditions are met. The subsequent
question of whether the two-brane setup can be stabilized is left
for a future numerical study.

\section{Tensor perturbations}

Having obtained the background solutions, we are now interested in
perturbing about them. In particular, we focus on tensor perturbations
in the brane-bulk system in order to understand the nature of gravity
for a brane-bound observer.

The most convenient framework in which to describe perturbations is
the explicitly conformal line-element\begin{equation}
ds^{2}=a^{2}(Y)\left(\eta_{\mu\nu}dx^{\mu}dx^{\nu}-dY^{2}\right)\label{eq:confmetric}\end{equation}
that follows from transforming line-element (\ref{eq:RSmetric}) to
a new coordinate: $Y=\int_{0}^{y}e^{-X(y')}dy'.$ The warp factor
is $a(Y)=e^{X(y)}.$ The gauge-invariant tensor perturbations decouple
from the scalar and vector perturbations, and take the form\begin{equation}
\delta g_{\mu\nu}^{(T)}=a^{2}h_{\mu\nu},\quad\delta g_{\mu5}^{(T)}=\delta g_{55}^{(T)}=0,\end{equation}
where indices on $h_{\mu\nu}$ are raised (lowered) by $\eta^{\mu\nu}$($\eta_{\mu\nu})$
and $h_{\mu\nu}$ is transverse traceless:\begin{equation}
h^{\mu}{}_{\mu}=\partial_{\mu}h^{\mu}{}_{\nu}=0.\end{equation}

The only wrinkle in this otherwise straightforward setup is that one
must also consider perturbation in the brane positions \cite{gt}.
To be precise, there is a scalar degree of freedom associated with
each brane, which adds an additional term to the perturbed junction
conditions. If there were only one brane \emph{and} no matter perturbation
on it, the perturbation in position could simply be eliminated by
a redefinition of the coordinates. In our scenario, we will suppose
there is no matter perturbation on the second brane and consider its
position fixed. On the other hand, we wish to calculate the gravitational
response to a point-mass on the first brane and thus its position,
here corresponding to the radion, must be perturbed.

\subsection{Equations of motion}

In terms of our new coordinate, the original unperturbed brane is
located $Y=0.$ We now suppose it is perturbed to $Y=\zeta(x^{\mu})$---to
first order, $y=\zeta(x^{\mu})$ also---following a general matter
perturbation given by the energy-momentum tensor $\tau_{\mu\nu}.$
The resulting equations of motion for the tensor modes are\begin{equation}
\left(\partial_{Y}^{2}+3\frac{a_{Y}}{a}\partial_{Y}-\Box\right)h_{\mu\nu}=0,\label{eq:eom}\end{equation}
subject to the junction conditions\begin{eqnarray}
\left.\partial_{Y}h_{\mu\nu}\right|_{Y\downarrow0} & = & \tau_{\mu\nu}-\myfrac13\eta_{\mu\nu}\tau-2\zeta_{,\mu\nu}\equiv\sigma_{\mu\nu},\label{eq:jc}\\
\left.\partial_{Y}h_{\mu\nu}\right|_{Y\uparrow Y_{\star}} & = & 0.\label{eq:jc2}\end{eqnarray}
Because there are two conditions, we expect only a discrete set of
gravitational modes. 

The fact that the left-hand side of (\ref{eq:jc}) is traceless gives
us the important equation for the displacement of the brane \cite{gt},
namely\begin{equation}
\Box\zeta=-\myfrac16\tau.\label{eq:bend}\end{equation}
If we assume $\partial_{\mu}\tau^{\mu}{}_{\nu}=0,$ then the transverse
nature of $h_{\mu\nu}$ is reflected in eq. (\ref{eq:bend}) as well.
In principle, we must solve this equation and substitute the solution
into eq. (\ref{eq:jc}). However, in this paper, we will content ourselves
with computing the appropriate Green's function from which $h_{\mu\nu}$
can be constructed.

\subsection{Green's function solution}

Equations (\ref{eq:eom}), (\ref{eq:jc}) and (\ref{eq:jc2}) can
be conveniently combined as\begin{equation}
\left(\partial_{Y}^{2}+3\frac{a_{Y}}{a}\partial_{Y}-\Box\right)h_{\mu\nu}=2\sigma_{\mu\nu}(x)\delta(Y),\end{equation}
allowing us to write a formal solution in terms of the retarded Green's
function given by\begin{equation}
\left(\partial_{Y}^{2}+3\frac{a_{Y}}{a}\partial_{Y}-\Box\right)G(x,x',Y,Y')=\delta^{(4)}(x\!-\! x')\delta(Y\!-\! Y').\end{equation}
After a Fourier transform, this becomes a 1-dimensional Green's function
problem\begin{equation}
\hat{D}_{k}\tilde{G}(Y,Y')\equiv\left(\partial_{Y}^{2}+3\frac{a_{Y}}{a}\partial_{Y}+k^{2}\right)\tilde{G}=\delta(Y\!-\! Y'),\end{equation}
which can be solved in terms of the complete set of normalized eigenfunctions
of $\hat{D}_{k}.$

To be precise, we suppose $\hat{D}_{k}\chi_{m}=(k^{2}-m^{2})\chi_{m},$
with boundary conditions $\partial_{Y}\chi_{m}(0)=\partial_{Y}\chi_{m}(Y_{\star})=0.$
(These boundary conditions reflect both the continuity of $\partial_{Y}\chi_{m}$
and the requirement of $\mbox{$\mathbb{Z}_{2}$}$-symmetry.) Then,
introducing $\psi_{m}=a^{3/2}\chi_{m},$ we obtain the Schr\"{o}dinger-like equation\begin{equation}
\left(\partial_{Y}^{2}+m^{2}-\frac{(a^{3/2})_{YY}}{a^{3/2}}\right)\psi_{m}=0.\label{eq:schrodinger}\end{equation}
This is usefully rewritten as \cite{lev}\begin{equation}
\left(\hat{D}_{+}\hat{D}_{-}+m^{2}\right)\psi_{m}=0,\label{eq:lev}\end{equation}
where\begin{equation}
\hat{D}_{\pm}=\partial_{Y}\pm\myfrac32X_{Y}.\end{equation}
The boundary conditions on the $\psi_{m}$ can be written as $\hat{D}_{-}\psi_{m}(0)=\hat{D}_{-}\psi_{m}(Y_{\star})=0,$
and the eigenfunction normalization is $2\int_{0}^{Y_{\star}}dY\,\psi_{m}\psi_{n}=\delta_{mn}.$
Then, \begin{equation}
\tilde{G}(Y,Y')=\sum_{m}a^{-3}\frac{\psi_{m}(Y)\psi_{m}(Y')}{k^{2}-m^{2}}.\end{equation}

The information we particularly want from the Green's function is
the response, in the stationary limit, at position $\mathbf{x}$ on
the brane due to a disturbance at $\mathbf{x}'$ also on the brane.
We have\begin{eqnarray}
G(\mathbf{x},\mathbf{x}') & = & \int dt'\int\frac{d^{4}k}{(2\pi)^{4}}\tilde{G}(0,0)e^{ik_{\mu}(x^{\mu}-x'^{\mu})}\nonumber \\
 & = & -\int\frac{d^{3}k}{(2\pi)^{3}}\sum_{m}\frac{\psi_{m}^{2}(0)}{\mathbf{k}^{2}+m^{2}}e^{-i\mathbf{k}\cdot(\mathbf{x}-\mathbf{x}')}\nonumber \\
 & = & -\frac{1}{4\pi r}\sum_{m}\psi_{m}^{2}(0)e^{-mr},\label{eq:greens}\end{eqnarray}
where $r=|\mathbf{x}-\mathbf{x}'|.$ A similar analysis was carried
out in \cite{callin}.

\subsection{Recovering the Newtonian limit}

Inspection of eq. (\ref{eq:greens}) shows we obtain the necessary
leading $1/r$-behavior if there is a zero mode. The contribution
of other modes may spoil this result if there are tachyon modes, or
if there are modes with $mr\ll1.$ This latter scenario is model-dependent
but we can show \cite{lev} that there is always a zero mode and never
a tachyon mode, i.e. $m^{2}\geq0.$

Multiplying eq. (\ref{eq:lev}) by $\psi_{m}$ and integrating over
$Y$ gives\begin{eqnarray}
m^{2} & = & -2\int_{0}^{Y_{\star}}dY\,\psi_{m}\hat{D}_{+}\hat{D}_{-}\psi_{m}\nonumber \\
 & = & -2\psi_{m}\hat{D}_{-}\psi_{m}\left.\right|_{0}^{Y_{\star}}+2\int_{0}^{Y_{\star}}dY\,(\hat{D}_{-}\psi_{m})^{2}.\end{eqnarray}
The first term is zero by the boundary conditions and the second term
is non-negative. Furthermore, the zero mode is given by $\hat{D}_{-}\psi_{0}=0,$
that is\begin{equation}
\psi_{0}\propto a^{3/2}.\end{equation}

\section{Gravity and a massive bulk scalar field}

We now apply the above formalism to the specific background solutions
computed in section \ref{sec:Background-solutions}. In both cases,
small $M^{2}$ and large $-M^{2},$ we will concentrate on the situation
at zeroth order in $\epsilon.$ For a start, as we have seen, nonlinear
perturbation theory already gives perturbative effects at zeroth order,
and secondly, because eq. (\ref{eq:schrodinger}) is of Schr\"{o}dinger-type, it
is straightforward, though hardly illuminating, to compute perturbative
corrections to the $\psi_{m}.$ We will assume that $Y_{\star}\ll r$
and call this the long-distance limit.

\subsection{Small $M^{2}$}

In this limit, we cannot completely ignore the first order term in
our solution for $X$ because it is necessary for getting the boundary
condition at $Y=0$ right. However, to a good approximation, we may
account for $X_{1}$ by simply rescaling $X_{0}:$\begin{equation}
X\simeq\myfrac14B\ln\left(1-\myfrac23B^{-1}y\right),\end{equation}
where $B=1-\frac{1}{6}\epsilon.$ It follows that\begin{equation}
a\simeq\left(1-\myfrac{Y}{Y_{s}}\right)^{Y_{s}/6},\end{equation}
with $Y_{s}=2-\frac{4}{9}\epsilon$ to first order in $\epsilon.$
The solutions to eq. (\ref{eq:schrodinger}) are then built from linear
combinations of Bessel functions of the first and second kind:\begin{equation}
\psi_{m}=(Y_{s}-Y)^{1/2}{\cal J}_{p}\left(m(Y_{s}-Y)\right),\end{equation}
where $p=\frac{1}{4}(Y_{s}-2)\simeq-\frac{1}{9}\epsilon.$ The boundary
conditions become ${\cal J{}}_{p+1}(mY_{s})={\cal J}_{p+1}(m(Y_{s}-Y_{\star}))=0.$
In the limit of large $mY_{s},$ we find\begin{equation}
\Delta m=\pi Y_{\star}^{-1}>\pi Y_{s}^{-1}\gg r^{-1},\end{equation}
and therefore that the infinite tower of states will not seriously
alter (\ref{eq:greens}). The Newtonian limit will also not be affected
by light modes: there is no massive bound state in the limit of small
$mY_{s},$ and we conclude that the lightest mode always satisfies\begin{equation}
m_{1}\sim Y_{s}^{-1}.\end{equation}

\subsection{Large $-M^{2}$}

We can use the zeroth order term for $X$ to compute the coordinate
$Y.$ However, the second order term $E_{1}(y(Y)/\epsilon)$ is crucial
to the computation of the effective potential in (\ref{eq:schrodinger}):
after being differentiated twice, it becomes a zeroth order quantity.
Actually, it is slightly more convenient to use the physical coordinate
$y.$ Letting $\xi_{m}=a^{1/2}\psi_{m},$ we have\begin{equation}
\left(\partial_{y}^{2}+\frac{m^{2}}{1-\frac{y}{3}}+\myfrac1{12}\frac{\cos\frac{2y}{\epsilon}}{(1-\frac{y}{3})^{2}}\right)\xi_{m}=0.\end{equation}
There is some hope of solving this equation in the limit that the
regulator brane is brought very close to the first brane. To be specific,
we suppose $y_{\star}\sim\epsilon.$ Then letting $y=\epsilon z,$
we obtain a Mathieu equation\begin{equation}
\left(\partial_{z}^{2}+A-2q\cos2z\right)\xi_{m}=0,\label{eq:mathieu}\end{equation}
where $A=\epsilon^{2}m^{2}$ and $q=-\frac{1}{24}\epsilon^{2},$ subject
to $\partial_{z}\xi_{m}(0)=\partial_{z}\xi_{m}(z_{\star})=-\frac{1}{3}\epsilon.$
If we could find bound states with $m\sim{\cal O}(\epsilon^{n}),\, n>-1,$
it might be that $my_{\star}\ll mr\ll1,$ and this would be fatal
for the Newtonian limit. As it happens, this does not occur. The solutions
\cite{aands} to (\ref{eq:mathieu}) are linear combinations of $F_{\nu}(\pm z),$
which for small $q$ become\begin{equation}
F_{\nu}(z)\simeq e^{i\nu z}\left[1-\myfrac14q\left(\frac{e^{2iz}}{\nu+1}-\frac{e^{-2iz}}{\nu-1}\right)\right],\end{equation}
where $A\simeq\nu^{2}+\frac{1}{2}q^{2}/(v^{2}-1)$ for $\nu$ real
and non-integer. Then, one can show that it is not possible to satisfy
the boundary conditions for small, real $\nu.$ This implies that
either $\nu\sim1$ or $\nu$ is purely imaginary, i.e., in an instability
band of the Mathieu equation. In either case, $A\sim1.$

\section{Conclusions}

We have conducted two distinct perturbative analyses of brane world
models that incorporate a massive bulk scalar field. First, we used
nonlinear perturbation theory to find background solutions in the
limits of small mass and large negative mass. In both cases, we found
it was necessary to add a second brane to eliminate a singularity.
Second, we considered tensor perturbations about these background
solutions. We were able to show that the brane-bound observer will
observe only slight modifications to Newtonian gravity in the case
of a light bulk scalar field. This result also holds true for a bulk
scalar field with large negative mass, which in some ways mimics a
bulk cosmological constant. Even though such a bulk potential may
be unphysical, it arises naturally in higher-derivative theories of
gravity that are conformally related to scalar-tensor theories, including
general relativity with scalar field matter. Thus our analysis sheds
light on the viability of these theories in the context of bulk extra
dimensions. We will make this connection more rigorous in upcoming
work \cite{pp}. We also look to study the stability of the two-brane
system introduced here.

\begin{acknowledgments}
It is a pleasure to thank Dorothea Deeg and Dani\`{e}le Steer for
illuminating discussions. MP was supported by SFB 375.
\end{acknowledgments}

\end{document}